# Letting the Brain Speak for Itself.

Gerhard Werner
Department of Biomedical Engineering,
University of Texas at Austin, TX

Abstract:

Metaphors of Computation and Information tended to detract attention from the intrinsic modes of neural system functions, uncontaminated by the observer's role in collection and interpretation of experimental data. Recognizing the self-referential mode of function, and the propensity for self-organization to critical states requires a fundamentally new orientation, based on Complex System Dynamics as non-ergodic, non-stationary processes with inverse power-law statistical distributions. Accordingly, local cooperative processes, intrinsic to neural structures and of fractal nature, call for applying Fractional Calculus and models of Random Walks with memory in Theoretical Neuroscience studies.

Introduction:

   Conceptual frames in which we ordinarily think and interact  are, in general,  fundamentally metaphorical in nature (Lakoff and  Johnson, 1980):  familiar patterns are sources of organizing the understanding of novel situations, offer convenient locutions for application to less defined or ill understood circumstances, and assist with selecting decisions and actions (Ortony, 1993). In Science, metaphors have traditionally influenced the formation of scientific concepts and theories, and supplied evocative terms for their formulation (Harre, 1995). They are also often credited with stimulating novel ideas and generalizations, and suggesting useful experimental approaches (Paton, 1996). Yet, caution is in order:  beware the bearers of false gifts!  The correspondence between a given situation or problem area, as the target,  with the source  of a metaphor thought to offer a suitable metaphoric relation, is in general based on surface appearance.  At a deeper level, the source may derive its validity from underlying assumptions and embedded conditions that diverge from, and may in fact conflict with, the conceptual structure of your target domain. You may find yourself now unwittingly applying the conceptually deep structure of your source to the target: you imported a deep structure for which you did not bargain. In Werner (2004), I illustrated errors and constraints that can thereby arise  for the interpretation of observations in certain neurophysiological experiments. In the least, commitment to the metaphor may lead one to overlooking or disregarding  more pertinent alternatives. In this spirit, Eliasmith (2003) called for  "moving beyond metaphors", citing symbolism, connectionism and dynamicism as three pervasive metaphoric domains of traditional System Neuroscience. In the following, I will at first briefly review my own list of metaphoric obstacles (partly divergent from Eliasmith), with emphasis on their historical origin, and the nature of constraints they have imposed on conceptualizing  System Neuroscience. I will then review what I believe we have learned in the last 20 years by recognizing the brain as a self-organizing  complex dynamical system in a state of criticality.

The Metaphorical Brain of Cybernetic Origin.

   This section's heading is also title of Arbib's (1972) seminal book which presented authoritatively the range and scope of Cybernetics' impact on thinking in the Neuroscience. Those old enough to have witnessed the rise and consolidation of Cybernetics in the decades of the 1940s and 1950s will recall the excitement, fervor of novelty, and the promise of new horizons for conceptualizing the nervous system. In a short span of time, Wiener's Cybernetics with the notions of feedback and control, Shannon's information theory, the concept of the Turing machine and the Church-Turing thesis, and von Neumann's invention of programmability of electronic computers appeared as the goldmines that would allow making sense of the activity of spiking neurons for the brain's information processing strategies, and ultimately suggest generalizations of societal import. For exploring the scientific and social implications of these innovations, the Macy Foundation sponsored the annual conferences of the "Cybernetics Group", beginning in 1943 and extending over 10 consecutive years. Composition of the group varied from year to year, ensuring the inclusion of a wide array of disciplines. A principal mover of these meetings was Warren McCulloch, to whom I will return later. The fascinating story of this group's wide ranging and often contentious deliberations is told in books by Heims (1991), Dupuy (1994) and in sections of Hayles' book of 1999. Transcripts of the proceedings were finally published by Pias in 2003.

   We owe the Metaphorical Brain two specific Metaphors: one concerning computation in and by the brain, the other considering the brain as an information processor. The next two sections will address them, in this order. While this essay is, sadly, in part an account of 'sic transit gloria mundi', it also aims to show the role of the Cybernetics movement in the dialectic to the next steps of conceptualizations.

   The Computation Metaphor

   This topic was reviewed in some detail at an earlier occasion (Werner, 2007) and I will therefore limit myself here to a few essential aspects that bear on the specific objective of this essay. There are two aspects to the computation metaphor: programmable Symbolic computation on Turing machines on the one hand, and Connectionism, on the other. The former lost its glamour for the practicing Neuroscientist fairly quickly: Turing machine computation (like any other form of programmable computation) transforms abstract objects by syntactic rules. Both, rules and semantics, must be supplied by the user. Hence, this form of computation is not part of a natural (i.e. user-observer independent) Ontology (Hayes et al, 1992). Moreover, almost all of physics is framed in Real Numbers which, due to computing errors by rounding and truncation, are not materially executable with the advertised accuracy (Landauer,1999 ; see also Carthright's (1983) "Why the laws of Physics lie". Most importantly, this form of Computationalism raised the uncomfortable question of representation: in Artificial Intelligence applications, and in many areas of Cognitive Science, the semantic (propositional) content is supplied by the user (Scheutz, 2002). The Neuroscientist, on the other hand, needs to account for a natural origin of the representations on which neural systems would perform computational transformations. An 'Internalist's Semantics' would be required for conferring semantic import to neural states, appealing exclusively to mechanisms internal to the brain (Grush, 2001). Failing, there may be a way around this : drawing on constructs from Control Theory and Signal Processing, Grush (2004,2009) suggests that the organism's sensori-motor engagement could deliver the functionality required for Computationalism to work. This seems of importance for generating Internal representation in Artificial Intelligence and Robotics for which Neurophysiology and Cognitive Theory are expected to provide useful heuristics (see for instance: Shanahan, 2005). But for the Neurophysiologist interested in the principles of brain function, representations pose a major hurdle: ultimately they must originate from the sensory signals the brain receives, whether by sensori-motor activity or simply by receiving sensory signals . Under the heading of 'Neural coding' , this subject will be addressed in the context of the Information Metaphor.

Neural Networks seemed to offer a new perspective on computation in the brain: Connectionism's appeal for the Neurophysiologist are due to its sharing -at least on a superficial level- some basic features of neural systems: a densely interconnected network of processing units ('neurons') that interact with one another by sending and receiving signals modulated by the weights associated with the connections between them. Two signal contributions have set tone and problematic of computation in these stylized Neural Networks (Hertz et al., 1991): considering simple model neurons as binary threshold units for computing weighted sums of their inputs, McCulloch and Pitts (1943) proved that an asynchronous assembly of such elements performs, in principle, Universal Computation (given certain choices of weights). Forty years later, Hopfield (1982) initiated the burgeoning era of Neural Network computation which, ultimately, rests on the definition of energy as a state function over a network of threshold elements; together, they display emergent collective computational abilities. Not only can networks of more realistic (e.g.: spiking) neurons of various formal properties realize any Turing computable function (Siegelmann and Sonnatg, 1995); under certain conditions, they can even outperform them (Siegelmann, 2003).

Broadly speaking, the thrust of the virtually incessant stream of publications on Neural Network Computation falls into two categories: one, to conceptualize a however tenuous connection to the Symbolic Computation paradigm. Smolensky's (1987) "on the proper treatment of Connectionism" is a valuable repository of the attempts to align symbolic and neural computation, introducing a subsymbolic paradigm as a kind of half-way measure: rather than hard syntactic rules, cooperation among "soft constraints" would collectively deliver inferences by a kind of parallel relaxation, conceivably emulating some features of cognitive processes. Cognitive Science has made extensive use of this principle's elaboration and extension (see, for instance: Feldman and Ballard, 1982). The other, neurobiologically more important category takes its directive from Hopfield' s original conceptual alignment with Statistical Mechanics: it stresses the cooperative behavior and emergent computational properties of connected networks of simple processors (e.g.: Sompolinsky,1981; Amit, 1989) , including their propensity for forming stable attractor states ( Rolls,2010), and attractor networks (Albantakis and Deco, 2011). The capacity of such networks for self-organization (Linsker, 1988; Hoshino et al, 1996), phase transitions (Opper and Kinzel, 1996; Kinzel 1997), and their natural interpretation as vector-to-vector transformers places the resources of dynamical systems at their disposal (Pellionisz and Llinas, 1982). Endowed with plastic synapses for changing synaptic weights, and capitalizing on the representational capacity of State Spaces, such vector spaces provide the required flexibility for multiple processing layers and recurrency. In this perspective, and at a higher level of abstraction, the entire pattern of the network's neuron activity is represented as a point in state space; activity patterns generated by an input vector that the network has learned to group together cluster to a circumscriptive cloud in the state space; and learning traces a trajectory in state space along the error dimension. Representations (concepts) can be portrayed as State Space partitions (Churchland, 1987, 1989, 1995).

The bulk of recent and current studies with neural networks emphasize a dual allegiance, true to their name: to neural (or neuron-like) elements on the one hand, and to network dynamics, on the other. Pursuing this latter avenue has uncovered surprising results. Consider, for instance that neural networks can self-organize to critical states (Bornholdt and Roehl, 2003; Levina et al, 2007 a) display avalanche dynamics ( Levina et al, 2007b); and how synaptic plasticity can drive self-organizing neural networks toward criticality (Meisel and Gross, 2009). In another series of studies, de Arcangelis and Herrmann

(2010) showed that avalanches formed in self-organizing neural networks can learn complex rules at phase transition, as result of a collective process. Vogels et al. (2005) direct attention to various forms of network dynamics and complex patterns of signal propagation with interrelations between stimulus driven and internally sustained network activity. These and similar phenomena violate the intent of the 'framers' of the Computation Metaphor for whom computation was discrete (and generally synchronous) in the programmable case, and continuous in Neural Networks. Aspects of nonlinear dynamics which are the reason for these violations will be discussed in later sections. They invite forcefully the timely question "Are biological systems poised at criticality" (Moran and Bialek, 2010) ? In any case, it appears amply justified to extricate Connectionism from its affiliation with the Computation Metaphor, and resurrect it in the perspective of "Computation on Networks", or "message-passing on graphs" (Mezard and Mora, 2009). However, as "old soldiers never die", it behooves one to be alert to the fact that the Computation Metaphor continues to influence, often subliminally, thought and speech patterns among practicing Neuroscientists. Viewing Connectionisms in the framework of the Computation Metaphor lets you forget its most distinctive features: propensity for discontinuous state transitions and self-organization. For the engineer, they are dreaded like the plague, and must be avoided. For the brain, they may the essential mode of operation. Accordingly, Connectionism in Neurophysiology should more appropriately viewed as a stage in the 'Dynamic Turn' of a later section in this essay.

The Information Metaphor

Information turned out to be among the most embattled and controversial topics in the discussions of the Cybernetics Group. As is well known, Shannon (1948) developed the Mathematical Theory of Communication (MTC) which delivers a quantitative measure of the accuracy (and correspondingly, of uncertainty) with which a message from a sender can be received by a receiver. Il is based on the selection by the receiver of one of the elements in a predefined ensemble. Shannon also established the formal equivalence of the uncertainty with physical Entropy in closed thermodynamic systems. Studying control system, Wiener (1948) arrived independently at the same mathematical result. In sometimes acrimonious deliberations of the Cybernetics Group, MacKay (1969), proposed to complement Shannon's 'selective information' with a 'structural information' which would capture the stepwise accretion of elements to a composite (like viewing a picture in a sequence of scans), and 'semantic information' as the selective operation which a signal performs on the recipient's set of possible states of behavioral readiness (McKay, 1954). For a more detailed discussion, see: Werner, 1989; Werner, 2007: section 2.3). However, Shannon carried the day in the Group's deliberations: the neat quantification of selective information was just too seductive to compete with "muddy" meaning and semantics (Adams, 2003).

The one aspect of these debates that continues to be particularly relevant for Neuroscience pertains to "Neural Coding": for the Computationalist of the programmable version, this was to be the source of internal representations; for the connectionist, it provided the input to the neural nets. As indicated before, activity in individual nerve fibers or neurons can be considered binary. It was then virtually irresistible to view the relation between input and output of a neuron in the framework of Shannon's information transmission from a sender to a receiver. This became what one may call the "hegemony of the digital doctrine" in Neuroscience. Interestingly and with a twist of irony, R. Gerard, the only Neurophysiologist of the Group, most strenuously objected to this notion, despite the fact that he and Li were in fact the first to record single neuron spikes from cortex. He argued that undue emphasis on patterns of single neuron activity would defeat appreciating the genuine nature of brain events.

Nevertheless, "Neural Coding" prevailed and has triggered a flood of theoretical and experimental studies. This field was most competently reviewed in 1997 by Rieke et al , but the stream of new investigations still persists incessantly.  In the foundational context of Information Theory, Neural Coding, Representation and Information Processing came to constitute a closely interrelated nexus of investigative targets (Borst and Theunissen, 1999). Two questions are directive: one, which feature of a neural spike train (rate, interval statistics, correlations, etc) carries (encodes) the message (in Shannon's sense)?; and second, how do downstream neurons 'evaluate' (decode) a putative message that may be encoded in a spike pattern  (see, e.g.: de Charms and Zador, 2000; Jacobs, et al., 2009) or in time-dependent signals (Bialek et al, 1991) ?.  Concerning the first issue, Perkel and Bullock (1968) listed 15 aspects of neural spike trains that could conceivably function as codes. Bullock (1993) returned to this topic 25 years later, claiming that many more variables may be contributing to neural integration at the system level: variables which, he thought, may be outside the then conventional neural modeling tools.

Failing to obtain in the short run any conclusive answer to both question, some investigators turned to an alternative approach: comparing  neural activity elicited by natural stimuli with known behavioral or psychophysical measures, the idea being that whatever measure of neural spike trains compares best with  corresponding perceptual -cognitive activity is then presumably a 'neural code'. For illustration of the basic pattern of this approach:  an early study of this type, involving cutaneous touch receptors and using firing rate as response measure, determined  that the spike count  (rate code) in peripheral efferent fibers suffices for reliably  distinguishing  8 different stimulus intensities (Werner and Mountcastle, 1965). This amounts to a capacity for transmitting 3 bits of information, which is also equal to the limit of cognitive processing in human subjects (Miller, 1956). For a review of numerous comparable studies examining correspondences between scales of neural activity and Psychophysics for different sense modalities, see Werner (1968). The same principle of seeking correlations between neural and perceptual cognitive activity became also the target of innumerable investigations with the 60 Hz cycles in sensory neurons, extending over several decades. For a recent summary and overview, see: von der Malsburg et al, (2010).

 What matters for the principal thesis of this essay is, however,  that  the views regarding the Coding Problem  diverged in time into two radically different directions: one, applying  mathematical approaches of increasing sophistication to  analysis, and generation, of spike trains in individual, or ensembles of,  neurons;  the other, calling the very notion of 'neural code' in question. One recent example of the former category is the elegant work of Haslinger et al (2009):  their approach determines a spike train's causal state model (i.e.a minimal hidden Markov model)  that generates time series which are statistically identical with the original spike trains. This enables a novel view to the coding issue, for it is then possible to relate the covariates to the causal states as generators of the spike train, rather than to the spike train itself. There was then also the question whether groups of discharging neurons might carry a message: for instance, Yule et al (2010) were interested in the information delivery rate from a population of neurons, with attention to redundancy of information within and between functional neuron classes. It also turned out that ongoing network states are effective determinants of an individual neuron's spiking activity (e.g: Harris et al, 2003; Shlens et al, 2006; Tang et al, 2008), and Truccolo et al (2010) reported that the ensemble history is a better predictor  for a neuron's spiking than is the ensemble's instantaneous state. These few recent and representative examples are indicative of the efforts to characterize statistical properties of spike trains, but they do not by themselves contribute to the question of message transfer on afferent pathways. However, the frequently applied measure of 'Mutual Information' is an effective way for determining the degree of independence of two data sets; it is independent of the underlying distributions and requires minimal assumptions about  dynamics and coupling of systems (Schreiber, 2000; Tkacic, 2010).

The second branch of the Coding Problem's history takes an entirely different view. Its origin can be traced to debates and controversies among the member of the original Cybernetics Group: the pivotal point was the contentious issue of the observer. As a paradigmatic situation, consider the usual experimental condition for studying "coding": typically, one applies stimuli whose metric one chooses as plausible, and then evaluates the neural (or behavioral) responses they evoked by a metric one chooses, such that one obtains a statistically valid (or otherwise to the experimenter meaningful) stimulus-response relationship. One then claims that the metric of the neural response 'encodes' the stimulus, being tempted to conclude that the organism is 'processing information' in the MTC paradigm. But recall that this paradigm deals with *selective* information; that is: the receiver needs to have available a known ensemble of stimuli from which to select the message. Moreover, this procedure does not permit one to know whether any of the metrics applied is of *intrinsic* significance to the organism under study: the observer made the choices on pragmatic grounds (Werner, 1988). In reality, once triggered by an external input, all that is accessible to the nervous system are the states of activity of its own neurons; hence it must be viewed as self-referring system.

   In a broader context, self-reference and self-organization became the rallying point of a successor to Cybernetics, generally known as "Second-order Cybernetics", with Heinz von Foerster (1981) as leading proponent. Closely related is Maturana's (1970) idea of Autopoiesis: in some ways a premonition of things that needed another 50 years to mature, as the Section on Nonanalytic Dynamics will show. In this view, and in contrast to the route of coding, stimuli from the periphery are thought to perturb the central nervous system's structure and internal organization according to its own internal dynamics (see also: Maturana and Varela, 1980). Coding and representing have in this framework lost legitimacy.

   The application of MTC in Neurophysiology is also vulnerable on other grounds. MTC and its generalizations to Information Theory and 'information processing' are predicated on the assumption of normalcy of data distributions, and ergodicity of the data generating process. But the abundant evidence for fractality and self-similarity at all levels of neural organization violates this assumption (Werner, 2010). More about this in the section on Nonanalytic Dynamics. For different reasons, the Neuronal Group Selection theory takes the view that the complexity, variability and unpredictability of the world precludes the notion of preexisting information, applicable to all situations, which the selective information paradigm of MTC requires (Edelman and Finkel,, 1984).

   At the time of its conception, MTC blended beautifully with the theory of programmable computation to forge an alliance that has molded an immensely influential Metaphor; the colloquial 'Information' of prevailing linguistic use undoubtedly fostering its ready acceptance. This entailed forgetting that Information (technically speaking) and Computation (of the Turing type) are observer constructs. Lacking an intrinsic ontology, Information refers to a *description* of reality, and Computation to an user-relative semantics. Nevertheless, bearing this in mind, both metaphors can serve useful purposes (Cox, 1946; Knuth, 2010). Information and computation metaphor are valuable constructs in the service of neurocomputational engineering developments (Eliasmith and Anderson, 2003; Eliasmith, 2009): in these applications, *You* are the master in your house: *You* design and build an object whose Ontology is therefore transparent to You, and to which *You* can choose the epistemic access to suit your purpose. But as Neurophysiologist, you face an unknown Ontology; the best you can do is to try finding an epistemic access that, by some criterion of your choice, is optimal: a kind of inverse problem. As I tried to show before: Metaphors are likely to be treacherous guides in this endeavor.

Analytic Brain Dynamics

   The starting point was W. Freeman's insight that  electrical activity recorded from aggregates of neurons (neuron masses,  in his originally terminology) can be interpreted as meaningful spatio-temporal patterns, related to sensory perception and learned behavior: from electroencephalographic records of the olfactory bulb in rabbits, he identified odor specific,  stable activity patterns which reflected  odor discriminations acquired by prior training. For each discriminated odorant, a learned limit cycle attractor is formed, differentiated from others by its basin and its spatial amplitude pattern. This ground-breaking  conceptualization  (and its multiple implications and extensions) was summarized in Freeman and Skarda (1985; Skarda and Freeman, 1987) as the result of some 12 years of work, and is also splendidly reviewed  by McKenna et al (1994) in the broader context of viewing  the brain as a physical system. This initial work introduced nonlinear dynamic analysis of neuronal system. Based on this work, Freeman confronted the Information Metaphor head on:  here are some excerpts from Freeman and Skarda (1985).: "context and meaning of representations are invariably in the brain of the observer and not of the observed. .... In the language of representations, the olfactory bulb extracts features, encodes information, recognizes stimulus patterns, …These are seductive phrases, ... but in animal physiology they are empty rhetoric."

   In the following years, Freeman's work continued in this vein with ever more sophisticated applications of non-linear dynamics to characterizing  records of human Electroencephalogram (EEG), which led to  identifying the role of chaotic phase transitions (Freeman, 2000;  Kozma and Freeman, 2002) and, eventually, to proposing that the non-linear brain dynamics is a  macroscopic manifestation  of a many-body field dynamics ( Freeman and Vitiello, 2006) . But this gets ahead of some developments in Physics with far reaching implications for virtually every field of knowledge, including, of course,  the Neurosciences.

   In the wake of the opening moves by Nicolis and Prigogine ( 1970)  and Haken (1975, 1977),  the study of non-equilibrium dynamic systems  in Physics yielded a massive body of knowledge and experiences that eventually came to constitute the modern theory of critical phenomena.  Stanley (1987, 1999) offers succinct accounts of the key concepts, as are the comments on theoretical foundations of particular relevance to neurobiology by Le Van Quyen (2003). Recall that a phase space is a mathematical construct: it is  understood as abstract space with independent coordinates representing the  dynamic variables needed to specify the instantaneous state of the system, which is represented as a point in that space. A system's state may be described at a microscopic level in terms of its constituent elements. It can also be characterized at a macroscopic level: that is how it appears to an observer who 'coarse grains' the microscopic state by lumping elements to larger aggregates. Nonlinear dynamic evolution equations determine the trajectory's motion in phase space. At certain points along the trajectory, singularities (discontinuities, bifurcations) of the evolution equations may be encountered which cause the system to undergo abruptly a phase transition to a qualitatively new and different state.  This may occur spontaneously, or in response to external perturbation. For applying this framework to Neuroscience, it must be understood that the differential equations for describing the state space dynamics stand in actual reality for the local physical and chemical processes that are in effect between neuronal elements; they are thought of as cooperative and collective interactions. In this spirit, the underlying neuronal dynamics is entirely internal to the modeled system, involving merely local collective processes among the constituting elements. If computational simulations adequately correspond to actual observations with real neural system, it is then thought that the dynamics of the real system is also sustained by system-intrinsic processes, safe of course for perturbations of external origin.

The plausibility of the dynamic view of brain function was forcefully enhanced by the discovery of spatially irregularly occurring patterns of propagated neuron discharges sequences in neural tissue: on the basis of detailed quantitative-statistical analysis, Beggs and Plenz, (1993, 1994) identified these patterns with the 'avalanches', characterized by Bak et al (1987, 1988) as evidence for a physical system having attained by self-organization a persistent critical state (see also Kadanoff, 1989). The principles underlying this claim are supported by numerous theoretical and experimental studies, and detailed aspects of the dynamics have been clarified by, for instance, Dickman et al (2000) and summarized in Sornette (2000). Criticality signifies here that minor perturbations, possibly spontaneous random noise, will trigger avalanches that correspond to phase transitions in the Theory of Critical Phase Transitions, referred to earlier. Plenz and Thiagarajan (2007) propose to view such avalanches as dynamic cell assemblies in neural tissue. Alerted by this discovery, numerous investigations of human fMRI under various conditions, summarized recently by Chialvo(2010) and Tagliazucchi and Chialvo (2011) solidly affirm the evidence for the type of complex emergent phenomena in brain that are typical of systems poised near to or at a critical state of (second order) phase transitions. Physics of condensed and excitable matter provides the theory of this phenomenon (e.g. Seth, 2006; Kadanoff, 1989), with the well-known Ising model serving as one of the physical prototypes: there, criticality corresponds to the emergence of correlation links to scale-free networks, with the same characteristic features that are observed with functional magnetic imaging in brains (Fraiman et al., 2009; Kitzbichler et al, 2009; Expert et al, 2010). Thus, the notion of brain criticality rests on relatively solid grounds, not as a Metaphor, but as the result of intrinsic physical mechanism. However, it is then also important to be aware that the mathematical tools for describing criticality in statistical systems is in general quite different from the language used when working with dynamical systems (Mora and Bialek, 2010). This will be pursued in the Section on nonanalytic Dynamics.

The implications of the dynamic conceptualization and criticality in brain physiology are discussed and illustrated in Werner (2007; 2009 a,b,c). Haken's approach, cited above, spawned the extensive investigations that eventually consolidated to the field of Coordination Dynamics (Haken et al, 1985; Kelso et al, 1992; Kelso, 1995 ). It established in different types of experiments and with careful analytic methods the concurrence of phase transitions in motor and perceptual-cognitive performance with brain electrical activity. In the course of these investigations, Kelso arrived at the view that the critical state in brain dynamics should be understood as "a space for exploring competing, perhaps conflicting, dynamic regimes" (Kelso and Engstroem, 2006). This may also bear on the proposal by Baylly and Longo (2011) that a kind of 'extended criticality' should be considered, possibly reflecting an entanglement of coexisting levels of order at the macroscopic. For the sake of completeness, recall that a 'dynamic turn' occurred also in Cognitive Science, where it was essentially set in motion by Port and van Gelder's (1995) book 'Mind in Motion'.

What is the significance of brain criticality? In the first place, it is associated with the establishment of long-range correlation for integration of activity across distant regions of neural tissue, displaying conspicuous fractal properties. The nature of this profound neuronal reorganization was ascertained under various experimental conditions and in numerous human EEG and fMRI studies ( for detailed citations, see : Turcotte, 1999; Werner, 2010): in this process, neurons assemble to new clusters whose size distribution scales with a power function of a negative exponent smaller than two, and exhibits self-similarity. Accordingly, the elements of the system assume a fractal order which is associated with entirely new properties: obeying, at the system's macroscopic level, new laws, and requiring new descriptors, which cannot be simply deduced from the prior state. We say then that a new ontology originates with phase transitions, which requires new epistemic criteria for its description and interpretation (Werner, 2011).

Here, then is the drastic difference to the cybernetic Information Metaphor: neural systems do not process information; rather, being perturbed by external events impinging on them, neural systems rearrange themselves by discontinuous phase transitions to new ontic states, formed by self-organization according to their internal dynamics. Internal to the system, these new ontic states are the 'raw material' for the ongoing system dynamics, by feedback from, or forward projection to other levels of organization. Looking at the dynamics from the outside, these dynamic states are objects to which the observer can do no more than apply his/her observer-dependent interpretation of accessible system observables. Clearly, if seen in this way, the metaphors of cybernetic origin and the brain dynamics sketched here are incommensurable: albeit interacting, inside and outside are two worlds, apart. How to deal with the (fractal) dialect the brain speaks in its own internal world is the subject of the next section.

<u>Brain Complexity with nonanalytic Dynamics</u>

Following this path will require delving into the burgeoning field of Complexity management. But first, a new look at recent findings of neuronal brain dynamics, based on EEG scalp records of normal subjects. Applying the time evolution of a minimum spanning tree method (Kruskal.1956) to analyzing human EEG, Bianco et al (2007) noted the intermittently occurring abrupt changes of topology, designated as "events' , which obeyed a renewal statistics. In the further course of the analysis, these authors concluded that EEG reflects a non-ergodic, non-Poisson, renewal process with a power law index < 2 (Bianco et al, 2008). Comparing this observation with results of an earlier study of the fluorescence intermittency in blinking quantum dots, Bianco et al, (2005) suggested that the recorded events can be attributed in both cases to a cooperative dynamics with emerging self-organized coupling of many interacting units (Grigolini, 2005; Bianco et al, 2008). In an independent series of experiments, Allegrini et al (2009) recorded coincidences of events, occurring simultaneously among two or more scalp electrodes. The waiting time distribution between consecutive events presented an inverse power-law index of approximately 2, corresponding to a perfect 1/f noise. These authors also proved that the coincidences are driven by a renewal process. Allegrini et al (2010) carried this line of research one step further by studying the Rapid Transition Processes (RTP) in EEG which Fingelkurts and Fingelkurts (2004) had examined in great detail. RTP's are not only evidence for intermittent global metastable transitions, but they also display multichannel avalanches (see Beggs and Plenz, l.c.) which appear as simultaneously occurring RTP's at several EEG recording sites. Statistical measures of multichannel avalanches exhibit inverse power law statistics and, thus, attest to the state of self-organized criticality of the entire cortex. Evidently, the units of activity underlying the recorded events are in these cases clusters of neurons (neuronal assemblies) whose dynamical interactions constitute networks.

Considering the case of two clusters of element (sub-networks) embedded in a network, Bianco et al (2008) found that cluster-to-cluster interaction is facilitated if the clusters are self-organized. Interactions occur then abruptly (designates as "crucial vents"), with the statistics of inter-event times indicating a non-ergodic, non-Poisson, renewal process (Turalska et al., 2009). West et al (2008) examined the conditions for maximizing information exchange between two complex networks, taking into account their characteristics as non-stationary processes with inverse-power law statistics. Viewing the power law exponent as measure of complexity, it turned out that information exchange is optimal when the complexities of perturbed and perturbing network are equal. Subsequently, it became apparent that perturbing and perturbed complex network must also share their respective temporal complexities, i.e. their nonstationary, nonergodic fluctuations at the onset of phase transitions (Turalska et al, 2011). Obviously, conditions for "information sharing' among complex systems are quite stringent, and much more work will be required to identify all relevant parameters. Surprisingly, complex systems do not respond to external stimuli at all unless they are not also complex (Turalska et al, 2009; Aquino et al, 2010).

In the Complex Network literature, it is customary to speak of "Information flow" or "Information exchange" in and between complex networks with the clear understanding that what is under discussion is NOT Shannon-type information, but rather Mutual Information, at best. My own preference is to view the relationship between complex networks as 'perturbation', with a perturbing network affecting topology and dynamics of a perturbed network, according to the latter's internal dynamics. Note the drastic difference between this outlook, and the notion of Information transmission between neurons (and neuronal systems) by encoded messages in the Cybernetic Information Metaphor.

An equally fundamental distinction is due to the complex webs being fractal: this renders ordinary or partial differential equations of motion no longer adequate for characterizing system dynamics (West et al, 2008). Instead, long-term memory in the dynamics of complex phenomena must be taken into account in the form of deterministic or stochastic fractional Dynamics. Fractional Gaussian noise was identified by Maxim et al (2005) in fMRI records of human brains, and Achard et al (2008) traced the fractal connectivity of long-memory networks. For other applications of fractional calculus in Neurophysiology, see for instance: Scafetta's et al (2009) work on postural control, and Lundstrom et al (2008) on fractional differentiation by neocortical pyramidal neurons. Suitable texts for this area of Mathematics are, for instance: West, Bologna, Grigolini (2003) and Miller and Ross (1993). Computational modeling on this basis may correspond in the Physics of Brain to Random Walks with long-term memory (West and Grigolini , 2011; Ch. 4; Montroll and Shlesinger,1984). Random Walks were used for tracing pair-wise local interactions along fractal neuronal connections by Sporns (2006), Fontoura Costa and Sporns (2006), and Fontoura Costa et al, (2011). Bieberich (2002) advocated the candidacy of the self-similar branching structure of recurrent fractal neural networks for the brain's local and global 'Information processing' (Sic !).

The observations and conclusions reported in this Section are in accord with the brain being in a critical state (Chialvo, 2008, 2010). In addition, they underscore the fractality of the critical state that results from the brain's self-organization. The brain's critical state can thus be viewed as a complex network of neuronal clusters on multiple scales: the clusters (neuronal assemblies) being the network nodes whose links enable coordinating their activity states. In virtue of the long-range connections between clusters, all parts of the system act in the critical state as if they can potentially communicate with each other, yet actual interactions are strictly local and constrained by seemingly stringent conditions of complexity matching.

Summary and Conclusion

Based on tracing notable landmarks in the history of Theoretical Neuroscience of the past six decades, evidence is presented that the metaphors of Computation and Information have stood in the way of gaining access to the intrinsic modes of neural system operation. Connectionism, originally subsumed under the Computation Metaphor, can more appropriately be viewed in the framework of Complex Networks. More importantly, Neural Dynamics must take into account the fractal nature of phase transitions and criticality in complex systems, requiring the mathematical tools of Fractional Calculus and its physical models of Random Walks with long-term memory: the overriding issue is the fractality in the context of Complex System Dynamics.  On account of this, self-similarity in neural organizations and dynamics poses one of the most intriguing and puzzling phenomenon, with potentially immense significance for efficient management of neural events on multiple spatial and temporal scales.   Answering the challenge posed in the title: "*fractal spoken here*".